# Key challenges for the surveillance of respiratory viruses: transitioning out of the acute phase of the SARS-CoV-2 pandemic

Oliver Eales[1,2,3], Michael J. Plank[4], Benjamin J. Cowling[5,6], Benjamin P. Howden[7,8], Adam J. Kucharski[9], Sheena G. Sullivan[3,10], Katelijn Vandemaele[11], Cecile Viboud[12], Steven Riley[1,2], James M. McCaw[3,13], Freya M. Shearer[3,*]

[1] School of Public Health, Imperial College London, UK

[2] MRC Centre for Global infectious Disease Analysis and Abdul Latif Jameel Institute for Disease and Emergency Analytics, Imperial College London, UK

[3] Centre for Epidemiology and Biostatistics, Melbourne School of Population and Global Health, The University of Melbourne, Australia

[4] School of Mathematics and Statistics, University of Canterbury, Christchurch, New Zealand

[5] Laboratory of Data Discovery for Health, Hong Kong Science and Technology Park, Shatin, Hong Kong Special Administrative Region, China

[6] WHO Collaborating Centre for Infectious Disease Epidemiology and Control, School of Public Health, LKS Faculty of Medicine, The University of Hong Kong, Hong Kong Special Administrative Region, China

[7] Microbiological Diagnostic Unit Public Health Laboratory, Department of Microbiology and Immunology, The University of Melbourne at the Peter Doherty Institute for Infection and Immunity, Melbourne, VIC, Australia

[8] Centre for Pathogen Genomics, University of Melbourne, Melbourne, Australia

[9] Centre for Epidemic Preparedness and Response, London School of Hygiene and Tropical Medicine, London, UK

[10] WHO Collaborating Centre for Reference and Research on Influenza, Royal Melbourne Hospital, and Department of Infectious Diseases, University of Melbourne, at the Peter Doherty Institute for Infection and Immunity, Melbourne, Vic, Australia

[11] Global Influenza Programme, World Health Organization, Geneva, Switzerland

[12] Division of International Epidemiology and Population Studies, Fogarty International Center, National Institutes of Health, Bethesda, MD, USA

[13] School of Mathematics and Statistics, The University of Melbourne, Australia

* Corresponding author: Freya Shearer freya.shearer@unimelb.edu.au.

**Abstract**

To support the ongoing management of viral respiratory diseases, many countries are moving towards an integrated model of surveillance for SARS-CoV-2, influenza, and other respiratory pathogens. While many surveillance approaches catalysed by the COVID-19 pandemic provide novel epidemiological insight, continuing them as implemented during the pandemic is unlikely to be feasible for non-emergency surveillance, and many have already been scaled back. Furthermore, given anticipated co-circulation of SARS-CoV-2 and influenza, surveillance activities in place prior to the pandemic require review and adjustment to ensure their ongoing value for public health. In this perspective, we highlight key challenges for the development of integrated models of surveillance. We discuss the relative strengths and limitations of different surveillance practices and studies, their contribution to epidemiological assessment, forecasting, and public health decision making.

## 1. Introduction

Surveillance plays a critical role in the management of epidemic diseases. This has most recently been demonstrated by the COVID-19 pandemic, during which existing approaches to respiratory pathogen surveillance, such as community testing, were rapidly scaled up and many enhanced or new surveillance activities, such as infection prevalence surveys, were implemented (1). The data generated by COVID-19 surveillance systems has provided situational awareness and informed myriad policy questions (2,3). The unprecedented circumstances of the pandemic have revealed both surveillance system strengths, where required data were available in a timely manner, and shortcomings, where data were delayed, unavailable or uninformative.

Many components of COVID-19 surveillance systems that were established to support pandemic response provided insight into epidemic dynamics and intervention impacts that were not possible through surveillance systems in place prior to the pandemic. For example, in the United Kingdom (UK) national infection prevalence surveys provided near-real-time insight into the *infection* (as opposed to case) dynamics of SARS-CoV-2 (4,5), including infection rates over time by age group (6). Household studies and the systematic collection of contact tracing data, particularly in Europe (7) and Asia (8), enabled estimation of key biological quantities affecting transmission (*e.g.*, the generation interval). Systematic collection of behavioural data from population surveys — with prominent examples in Australia (9), Hong Kong (10), and the UK (11) — provided information on the impact of public health measures, including the functioning of various surveillance components. International and country-level platforms for genomic data integration and reporting enabled early characterisation of variants of concern and provided insight into the global patterns of variant spread (12–15).

As we transition out of the pandemic, many countries have discontinued or scaled back COVID-19 surveillance activities and are moving towards an integrated model of surveillance for COVID-19, influenza, and other viral respiratory pathogens of epidemic or pandemic potential (16). Globally, COVID-19 continues to place a significant burden on population health and health systems (17). While seasonal patterns are not yet possible to predict, future epidemic waves of SARS-CoV-2 are anticipated (18), and we continue to face the threat of novel variants. With less social disruption due to COVID-19, seasonal circulation patterns of other respiratory viruses have resumed (19). The long-term impact of co-circulation and behavioural- and immune-mediated interactions between SARS-CoV-2 and other respiratory pathogens is unknown but is expected to place substantial additional pressure on healthcare systems. Effective surveillance will enable rapid understanding of the status of concurrent epidemics and preparation for any increased impact on healthcare during interpandemic periods. To assess, anticipate, and respond to this overall viral respiratory disease burden, the individual dynamics of each pathogen need to be monitored. Furthermore, effective surveillance will enhance levels of preparedness for responding to the next (inevitable) global respiratory virus emergency, including establishing criteria for activating (and deactivating) enhanced surveillance activities (such as special studies), according to surveillance objectives. While surveillance objectives will vary by locality and epidemiological context (as explored by the World Health Organization's Mosaic framework (20,21)) a collaborative surveillance approach will support national, regional and global preparedness (20,21).

To inform the design of a sustainable, integrated model of viral respiratory pathogen surveillance (Figure 1), now is a critical time to review both surveillance practices in place prior to the pandemic and novel approaches adopted for pandemic response. Resuming pathogen specific surveillance approaches, such as those for monitoring influenza, would represent a missed opportunity to build on learnings from emergency response efforts. Herein we describe key challenges for the design of integrated viral respiratory pathogen surveillance as the world transitions out of the COVID-19 pandemic. We focus on both challenges that apply to the monitoring of any individual viral respiratory pathogen and those that arise due to their co-circulation and integrated nature of a surveillance system. Furthermore, we explore how additional surveillance methods and studies can enhance knowledge and support our ability to anticipate the impact of these respiratory viruses

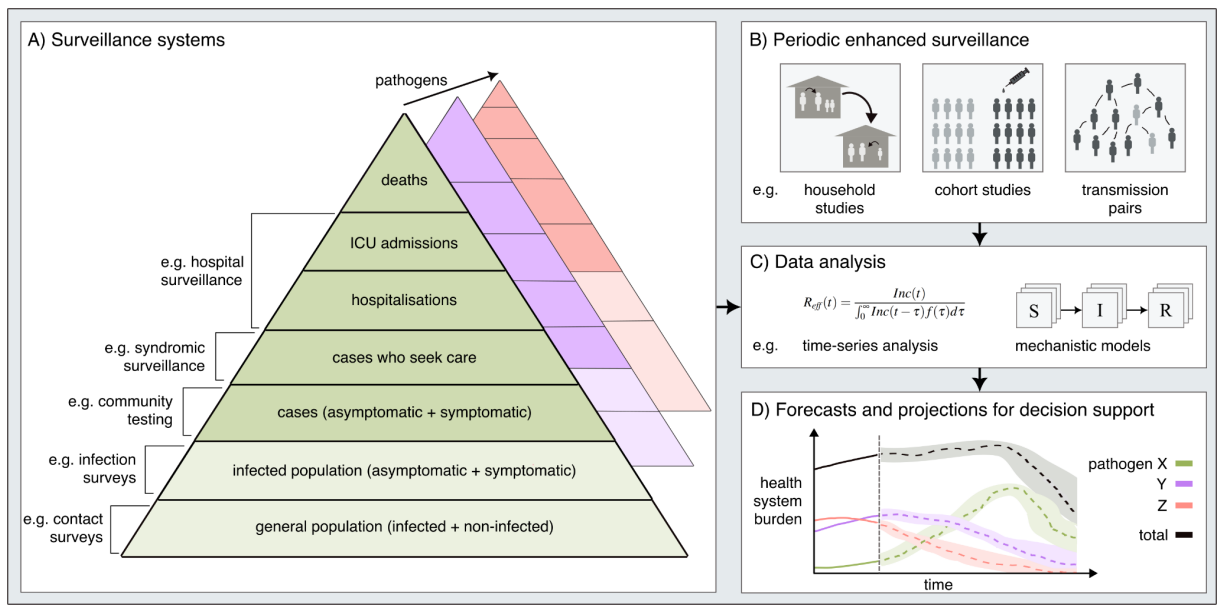

**Figure 1. An integrated surveillance system for multiple circulating respiratory pathogens.** (A) Surveillance systems, which collect data continuously, monitor changes in infection outcomes (or behaviour) at different levels of severity on the 'severity pyramid'. In general, the proportions in each level can vary over time (for example, due to the emergence of new variants) and so surveillance should be undertaken at multiple levels. Collecting additional behavioural data can help disentangle changes in the data streams due to behaviour specifically (e.g., changes in healthcare-seeking behaviour) or due to behaviour indirectly (e.g., changes in transmission rates due to changes in contact rates). In general, multiple pathogens will circulate concurrently, and it may be necessary for surveillance systems to be extended to distinguish between the different pathogens (see Section 2.1). Integrating genomic sequencing of samples collected at all levels can additionally allow differences in dynamics between variants to be identified. (B) Enhanced surveillance activities conducted periodically can augment continuous surveillance. Enhanced surveillance activities may inform on key epidemiological quantities, such as the generation interval and vaccine effectiveness, which are important for predicting future transmission dynamics. (C) Data collected from different sources of clinical, epidemiological, and genomic surveillance can be synthesised through data analysis pipelines and used to generate epidemic forecasts and scenario projections (long- and short-term). (D) These forecasts and projections can be used to support public health decision making and planning.

## 2. Measuring the clinical burden of multiple circulating respiratory pathogens

*2.1 How can sentinel syndromic surveillance handle multiple pathogens with overlapping symptom profiles?*

In many parts of the world, the public health response to COVID-19 — including stay-at-home orders (22,23) and a requirement to isolate if testing positive — strongly suppressed viral respiratory transmission (9). Accordingly, co-circulation of other respiratory pathogens such as influenza and respiratory syncytial virus (RSV) was limited (24,25). In the absence of other circulating pathogens, for individuals reporting symptoms, it could be assumed that SARS-CoV-2 was the causative virus, simplifying interpretation of syndromic surveillance data. In the context of co-circulating pathogens, interpretation of data from syndromic surveillance is more complicated; the overall clinical burden will most often be the result of multiple concurrent epidemics. As a case in point, the 2022 autumn season in the US was characterised by large overlapping epidemics of influenza, RSV and SARS-CoV-2, exerting unusual pressure on ICU capacity — a phenomenon coined the "tripledemic" (26).

Prior to the COVID-19 pandemic, surveillance of seasonal influenza primarily relied on 'sentinel syndromic surveillance', where a fraction of healthcare sites routinely report the daily number of individuals presenting with influenza-like illness (ILI) (27,28). ILI defining symptoms are not influenza specific, and the temporal signal can be biased by other respiratory pathogens with similar symptom profiles (29,30); for this reason, a subset of ILI cases undergo testing to estimate the proportion for which influenza was the causative virus (ILI+). The overlapping symptom profiles of COVID-19 and influenza will likely result in an increase in the number of individuals presenting with ILI (*i.e.*, a decrease in the specificity of ILI defining symptoms for identifying influenza infection) and previously suitable sampling strategies for testing may no longer yield a sufficient sample size for reliable evaluation of each virus's contribution to overall burden.

Expanding influenza sentinel syndromic surveillance to incorporate COVID-19 surveillance, as proposed by the World Health Organization (16), would avoid the need for two partially redundant surveillance systems. Just as for influenza surveillance, commonly used ILI definitions used for reporting would not (nor would be required to) capture all symptomatic COVID-19 infections (31). Broadening the symptom criteria (and increasing the total sample) for reporting would capture a greater fraction of symptomatic infections, irrespective of the underlying virus. However, since this could reduce the specificity of the symptom criteria, consideration should be given into how the symptom criteria might be optimised to provide the most epidemiological insight, while maintaining cost-effective sampling rates (low specificity would require greater sampling rates). As any syndromic signal detected would include infections due to influenza, SARS-CoV-2, and other pathogens, a random subset of individuals meeting the symptom criterion could undergo testing to identify potential causative virus(es) (32). This would enable the time-series for each circulating pathogen to be resolved, improving prediction of individual and overall disease burden (30) (see Figure 1D).

*2.2 How can trends in community levels of infection be monitored?*

Sentinel syndromic surveillance of ILI and ILI+ are typically assumed to be proportional to community levels of infection (33), despite these indicators only capturing infected individuals who report to a healthcare provider. As was already clear prior to COVID-19, the subset of the infected population seeking healthcare will be a biased subsample of the total infected population, and there is evidence of significant inter- and intra-season variation in the fraction of all infections detected for various reasons (34). Hence, developing systems for monitoring the underlying levels of infection in the general population is important in order to provide a less biased, more stable denominator for assessments of clinical severity (see section 3.2) and intervention effectiveness.

Measuring the true levels of infection is difficult and would require testing a representative sample of the general population regularly (see section 2.3). However, if quantities that reliably correlate with the population level of infection can be estimated (and their biases understood), the underlying dynamics of transmission can be inferred. This is not straightforward. First and foremost, case time-series data will not necessarily provide such a correlate. During the COVID-19 pandemic, many surveillance activities — including wastewater surveillance, symptom-based participatory studies, and mass testing — were used as correlates of SARS-CoV-2 infection, but not all of them will be sustainable into the future, and many have already been scaled back. Furthermore, while measuring correlates of infection rates is useful for estimating how infection rates vary over time i.e. relative changes, it does not provide estimates of the absolute level of infection in the population over time.

Wastewater surveillance of SARS-CoV-2 was used extensively during the COVID-19 pandemic (35–37). Monitoring viral concentration in wastewater over time could provide an indicator of relative changes in the level of incidence that is insensitive to changes in case ascertainment; when genomic sequencing is also performed on wastewater samples, changes in the level of incidence can be inferred for individual variants (38). However, to inform estimates of absolute infection levels, the relationship between viral concentration and incidence of infections must be robustly quantified, which while many approaches have been proposed (39–41), remains an open scientific challenge. Additionally, wastewater surveillance cannot identify who in a population is infected or provide any clinical information about the severity of those infections, and it is not yet clear if such approaches would be as efficient for influenza surveillance, with limited studies (42,43).

Symptom-based participatory surveillance — such as Outbreaks Near Me (44), FluTracking (45) and InfluenzaNet (46) — in which cohorts of individuals regularly report their symptoms, is a long-established approach to monitoring seasonal influenza (47). During the COVID-19 pandemic, those systems were rapidly adapted (or established) to contribute to SARS-CoV-2 surveillance (48,49). Now, with anticipated co-circulation of pathogens with similar symptom profiles, extending these surveillance approaches to include a respiratory sample testing component would be beneficial (50). At-home testing with validated, self-administered rapid antigen tests could provide a convenient option. However, symptom-based surveillance systems are sensitive to changes in symptomatology (51) and symptom reporting, and other surveillance systems that assess the full spectrum of disease caused by infection, including asymptomatic infection, would be required to monitor such changes.

At the outset of the COVID-19 pandemic, many countries implemented large-scale community testing to detect active infections, irrespective of symptoms or epidemiological risk (52). While not necessarily articulated in guidelines or surveillance plans, the primary objective of such mass testing was to support case and contact management to reduce transmission. The data generated by these programs (case time-series data) includes information on individuals who sought a test because they exhibited symptoms, were identified as contacts of confirmed cases, and/or were part of a workplace screening program (irrespective of symptoms), among other reasons. The fraction of infections detected by mass testing programs will not be representative of the infected population in ways that are difficult to quantify due to the range of reasons for testing. Furthermore, these biases will change over time (34) in response to policy, epidemiological, and socio-cultural factors.

As already described, all the above methods are highly useful for monitoring changes in infection prevalence, but they only measure correlates of infection. To estimate the infection prevalence from these correlates, we need to know how to map between infections and the correlates. This mapping will often depend on biological and/or behavioural factors (that need to be quantified) that should be expected to vary over time. An alternative approach to estimate the level of infection is to measure it directly. Random population testing, discussed below, is a key methodology that has been used to measure infection prevalence of SARS-CoV-2 and has been proposed for the ongoing surveillance of SARS-CoV-2 and other respiratory pathogens (53).

*2.3 What is the value added from random sampling studies for monitoring infections?*

By regularly testing a random subset of the population for the presence of a virus, one can obtain estimates of the infection prevalence over time. These studies are specifically designed to generate less biased estimates of infection prevalence compared to other testing strategies, and if demographic data are also collected, then heterogeneities in infection prevalence can be identified, and biases in the sample can be better understood. In contrast to most surveillance approaches that focus on symptomatic infections, random sampling studies identify asymptomatic and paucisymptomatic infections. Many countries have performed one-off small-scale studies to measure infection prevalence at a single point in time (54–56). However, the use of repeated random sampling for measuring infection prevalence through time has been extremely limited, with only two studies that we are aware of, both of which estimated infection prevalence of SARS-CoV-2 within the United Kingdom: the Office for National Statistics' (ONS) Coronavirus Infection Survey (CIS) (5) and the REal-time Assessment of Community Transmission - 1 (REACT-1) study (57). These studies enable robust estimation of the time-series of past infections, and in turn the proportion of the population who are no longer susceptible to infection due to recent exposure, improving epidemic forecasts and scenario projections (58). They can be extended (or integrated with other data sources) to gain greater insight into key epidemiological quantities, including real-world vaccine effectiveness (59,60), the transmission advantage of new variants (61,62), symptomatology (31), and post-viral sequelae (63). Additionally, indirect (mathematical and statistical) models based on other surveillance systems can be validated or refined, by quantifying their relationship to directly measured infection levels (64). Insights gained from suitably designed and regularly deployed infection prevalence studies are even greater when infection levels are high, and many positive individuals are identified.

Random sampling studies could be implemented (or continued) for the ongoing surveillance of SARS-CoV-2 and other respiratory pathogens, enabling less biased estimates of infection in the community, and supporting situational assessments and forward planning. However, this would not be without challenges. Over the course of the COVID-19 pandemic, there was a decline in recruitment rates in both the ONS CIS and REACT-1 studies – this has the potential to increase bias in the sample. Sustainability and feasibility both require further assessment since infection prevalence studies require significant investment. Different approaches, such as embedding randomised testing in existing cohorts or developing new indirect methods to assess infection levels, should be investigated, including the development of new analytical techniques to interpret the data. These may be trialled and evaluated in the seasonal context over the next 2-5 years in order to determine the most sustainable, informative and cost-effective ways to generate time-series of infection rates as part of routine viral respiratory surveillance. Testing for multiple pathogens in each sample instead of testing exclusively for SARS-CoV-2 would allow the infection prevalence and co-circulation of many respiratory pathogens (and their interactions) to be quantified; this is already being trialled by the ONS CIS (65). Implementing longitudinal studies (either separate or embedded) that employ frequent testing to quantify the relationship between the time since infection and test-positivity (66,67), could additionally allow the daily incidence of new infections to be estimated from estimates of infection prevalence. Further, long-term longitudinal studies could help quantify the risk of reinfection, and the degree of cross-immunity between different variants in the face of viral evolution.

## 3. Assessing changes in key biological quantities

*3.1 How can key epidemiological quantities be inferred?*

Surveillance approaches that collect information on pairs of "infectors" and "infectees" in successive chains of transmission are required for directly estimating (and monitoring for changes in) a number of key biological quantities that drive epidemic dynamics, including the incubation period, generation interval and relative contagiousness of asymptomatic infecteds (68). For example, the generation interval can most directly be estimated by linking dates of infection onset for infector–infectee pairs, data which are not collected through traditional "case-based" surveillance. While contact tracing activities and outbreak investigations do typically collect relevant data, they have not traditionally been conducted with that goal in mind, and may be scaled back or stopped entirely based on a perceived lack of need, leaving a potential knowledge gap. Knowledge of the generation interval is required for accurate estimation of the effective reproduction number (69), itself critical in evaluating the likely impact of alternative control strategies. To date, estimates of the generation interval for COVID-19 have relied on (often opportunistically) collected contact tracing data (8) (noting that these studies typically use symptom onset dates rather than infection onset dates, and so estimate the serial interval, which is then used as a proxy for the generation interval). While these data provide reliable estimates, they are only available when contact tracing is occurring, which is generally not the case for respiratory pathogens. If studies were specifically designed for this purpose, and/or embedded within other studies (*e.g.*, infection prevalence surveys) then data availability would not depend on the presence of contact tracing and or outbreak investigations. Data quality would likely be enhanced since we might expect more precise recording of time intervals of likely infection and symptom onset.

Further, the generation interval of SARS-CoV-2 changed as the pathogen evolved (70,71) and under different policy settings (8), and transmission settings (70). If surveillance were augmented by genomic data (see section 3.3) from infector-infectee pairs during periods of variant emergence, changes in the generation interval between variants could be measured in real-time, and used to assess potential mechanisms of transmission advantage. This would also allow estimation of the effective reproduction numbers of circulating variants during periods of variant replacement (72), improve real-time forecasts of variant impact, and inform intervention planning. Genomic data may also improve accuracy of the inferred transmission links e.g., excluding coincidental infections in close contacts that were actually acquired from independent sources.

*3.2 How can a virus' clinical severity be quantified?*

To predict the future numbers of severe outcomes (e.g., death, hospitalisation, etc.), the relationship between infection and the probability of a severe outcome needs to be well characterised, including how it varies with age, vaccination-status, comorbidities, and other epidemiological covariates. For this to be possible, accurate estimates of infection levels (to compare to the levels of severe outcomes) (73), or the infection outcomes of an unbiased sample of infected individuals are required.

The infection fatality ratio (IFR) and infection hospitalisation ratio (IHR) are measures of the proportion of infections that result in death or hospitalisation respectively. Over the course of the COVID-19 pandemic, the IFR and IHR for SARS-CoV-2 changed due to mass vaccination and the emergence of new variants (73). For seasonal influenza, in countries with suitable data systems and resources, the case fatality ratio (CFR) and case hospitalisation ratio (CHR) are routinely computed and may vary from season to season (74). However, it is difficult to distinguish the extent to which changes in the CFR and CHR for seasonal influenza are due to changes in case ascertainment or the intrinsic biology of the circulating strain and effectiveness of the vaccine. The IFR and IHR for seasonal influenza (which are independent of case ascertainment) are seldom known since the underlying levels of infection in the community are rarely measured and hospitalisations/deaths are not always attributed to a specific pathogen, undercounting the true burden. It is crucial we are able to measure any changes in the IFR and IHR of a pathogen so they can be incorporated into situational assessments and scenario planning (58). Severe outcomes are recorded by public health systems in many countries, but their definitions may vary between (and within) countries or over time and are often not universal, but sentinel based or for a subset of hospitals. And for mortality surveillance, this is often only available near real time for all cause mortality rather than for disease specific mortality. Linking the data collected on severe outcomes (and key covariates such as co-morbidities) to surveillance systems that identify the level of infection in the general population, could allow estimates of the IFR, and IHR to be made quickly. If the definitions used for severe outcomes were standardised through time, then changes in severity between and within epidemic seasons could be more easily detected.

*3.3 How can the emergence of antigenically distinct variants be accounted for when quantifying epidemiological quantities?*

High levels of population immunity drive a significant transmission advantage for antigenically distinct variants. Although emerging variants are often genetically similar to previously circulating variants, even small numbers of changes in critical regions of the genome can result in significant differences in their intrinsic biology and their resulting epidemic dynamics. Unprecedented rates of genomic sequencing (in some countries) during the COVID-19 pandemic (75) enabled epidemiological studies to distinguish infections caused by different SARS-CoV-2 variants, identify the eradication of outbreak clusters related to public health interventions (76), and identify differences in their epidemiological characteristics (62,70). However, in many countries, sequencing was either too limited or sampling strategies were not sufficiently effective to identify changes in the transmission dynamics due to the emergence of biologically distinct variants. Past surveillance of seasonal influenza has also been relatively limited in its (per capita or per case) rate of sequencing (77). More strategic integration of genomic sequencing (even at low levels) into future surveillance systems is required to capture epidemiologically relevant differences between variants (78–80).

Monitoring the mix of genetic variants circulating in the general community is critical for situational awareness, but in some countries the focus has been on monitoring SARS-CoV-2 variants in clinical settings (78). While sequencing of cases with severe disease provides important data on disease severity, including for the detection of changes in the IFR (see section 3.2), detecting if a variant has a transmission advantage should also be a priority, and this requires sequencing of cases from the general community (*e.g.*, embedded within an infection prevalence study (81)). An increase in inherent transmissibility of a pathogen (*i.e.*, increased basic reproduction number) causes clinical outcomes to grow exponentially (72) whereas the same proportional increase in the IFR only causes a linear increase in clinical outcomes. Of course from epidemic theory, over the longer term, for a highly transmissible pathogen, a transmission advantage will have only minimal effect on average prevalence (and the clinical burden), whereas an increase in the IFR will result in a linear increase in the clinical burden. By developing sequencing strategies that incorporate representative sampling of infections in the general population (e.g., through infection prevalence studies), variant proportions can be reliably estimated in near-real-time; and in the early stage of variant emergence, these data can be used to track transmission advantage for incorporation into epidemic forecasts and other situational analyses (72,82).

While the early pandemic was characterised by waves of distinct variants, the situation for SARS-CoV-2 now is that a diverse set of Omicron sub-variants are co-circulating at any time, with many differing by only a small number of mutations. It is important that a classification method exists so that this diverse strain set can be clustered into a smaller number of functional (from a public health response perspective) groupings with similar epidemiological characteristics; differences in epidemiological quantities can then be inferred between variants. During the pandemic, classification systems mainly grouped SARS-CoV-2 strains based on how closely they were phylogenetically related (83). However, convergent and directional evolution may lead to many overlapping clusters, or numerous genetically distinct strains which are antigenically similar (84). Antigenic cartography, which has been used extensively for influenza (85,86), could allow genetically diverse sets of strains to be clustered based on their antigenic characteristics (87); though there are some practicality issues for high throughput surveillance whilst SARS-CoV-2 is classified as a BSL3 virus.

### 4. Quantifying changes in epidemiological dynamics

*4.1 In an increasingly complex immune landscape how can infection history be quantified?*

Population immunity to SARS-CoV-2 infection has increased in complexity through the course of the pandemic. At the beginning of the pandemic there was likely minimal population immunity. With multiple vaccination courses, the emergence and spread of multiple variants, and waning of both vaccine- and infection-induced immunity, the individual- and population-level characteristics of immunity are now highly complex (88). Repeated epidemics of Omicron sub-variants indicate that immunity against reinfection may not be long-lasting, but reduced severity of infections over time is consistent with increasing levels of population immunity against severe disease (89). In principle all possible combinations of exposures and vaccinations, and their timing, must be considered to estimate immunity and the relative level of susceptibility. With future variants and updated vaccines, the problem will undergo a combinatorial explosion; this is a known challenge for influenza and there exist some modelling approaches designed to manage this complexity (90).

Significant improvements (for both influenza and SARS-CoV-2) could be made to forward projections of clinical burden and the design of future vaccination campaigns if the immune landscape could be more accurately quantified. Serological data, which is used to measure the presence of specific antibodies in an individual's blood, can be used to quantify immunity when the antibodies correlate sufficiently with protection against infection (91). Serological studies of influenza have greatly improved the understanding of long-term dynamics of influenza antibody-mediated immunity (92), but such serological data has not been used for routine surveillance. The collection of serological data was commonplace during the SARS-CoV-2 pandemic (93,94). Studies found that serum antibody titres correlated with protection against symptomatic infection during the first two years of the pandemic (95,96), with neutralising antibodies offering the strongest correlation (97,98). Most of the global population will now have antibodies against SARS-CoV-2 in their sera (following infection and/or vaccination), but future infections are likely to be caused by variants able to evade these existing antibodies. Additionally, the long-term dynamics of immunity against SARS-CoV-2 are not yet understood; correlate-of-protection studies (*e.g.*, longitudinal cohort studies) should be established to identify appropriate immunological tests that can quantify population- or individual-level immunity against infection with future SARS-CoV-2 variants. If suitable correlates-of-protection could be identified, then it is possible that serological surveillance systems could be established to quantify population immunity against newly emerging variants before waves of infection occur. If these immunological observations could be combined with predictions of strain evolution (99,100) the future burden of respiratory viruses could be anticipated.

*4.2 How will we determine the drivers of changes in future transmission dynamics?*

The COVID-19 pandemic saw marked heterogeneities in age-specific transmission, infection and severity of infection (101,102). As we transition out of the acute phase of the pandemic, the age distribution of infections will likely change due to changing patterns of population immunity (infection- and vaccine-induced) (103,104). Monitoring age-specific infection rates and transmission could improve epidemic projections by allowing differences between age-groups to be incorporated in epidemic models; this is highly important for projections of the clinical burden, since disease severity can be highly dependent on age (105–107). Further, comparison of age-specific infection rates with severe outcomes could allow any changes in the relationship between age and IFR to be identified and incorporated into situational assessments. Infection prevalence studies (see section 2.3) could explicitly measure differences in infection rates between age-groups, with fewer sampling biases than other testing strategies.

The future dynamics of SARS-CoV-2 transmission will not only be influenced by biological factors such as immunity. They will also be highly dependent on human behaviour, including changes in human contact networks, and changes in behaviour that influence the likelihood of transmission given contact. The pandemic saw dramatic changes in social contact networks (11), which altered the transmission dynamics of SARS-CoV-2 (108) and many other pathogens (109). With most travel and social restrictions now relaxed, levels of social mixing and movement have rebounded, but it is not clear when or whether contact networks will eventually reconfigure to their pre-pandemic patterns. Distinguishing changes in epidemiological dynamics due to changes in the biology of the virus from changes in human contact networks will be crucial for quantifying the future transmission dynamics of SARS-CoV-2 and other respiratory pathogens. Behavioural studies, which quantify human contact networks, and how they vary between populations and over time, will be paramount for identifying the origins of any epidemiological changes. Such studies will also support the refinement of epidemiological models that incorporate human behaviour, improving their value for situational assessment, planning and control.

*4.3 Can the frequency and timing of recurrent epidemics be projected?*

Long-term projections of the potential magnitude and timing of future epidemic seasons can assist health systems in better preparing for the resulting clinical burden. This work is distinct (and complementary) to the statistical near-term forecasting of peak timing and size routinely conducted throughout an epidemic season (110).

Influenza transmission is highly seasonal; in temperate regions there are yearly winter epidemics (111), whereas in tropical regions there is a high background level of influenza infection with additional epidemics occurring with less regular timings (112). The long-term dynamics of SARS-CoV-2 are uncertain, but recurrent epidemics are likely to occur into the future (113,114). In situations with significant circulation of SARS-CoV-2, influenza, and other respiratory pathogens it is not yet known how concurrent epidemics will interact. Virus-virus interference at the individual host level is a well-established phenomena (115–117), but far less is understood about the epidemiological consequences of that interference (111). Will concurrent epidemics cause increased pressure on health services? Or could epidemics interfere with each other, resulting in misalignment, or less predictable behaviour? Multiple recurrent epidemics involving SARS-CoV-2 and other established viruses will need to be observed to answer these questions. In the meantime, surveillance planning should consider these longer-term surveillance objectives. It is important that any data collected from surveillance systems is consistent between years, so that dynamics between years can be reliably compared and such questions over inter-epidemic timescales can be answered. If serological data were routinely collected, it might be possible to regularly estimate population immunity (see section 4.1), informing estimates of epidemic timing. Studies would first be required to identify when and what serological data should be collected, and by how much it could improve estimates of epidemic timing.

**5. Leveraging the global dynamics of transmission**

*5.1 How should data from across regions be incorporated into local situational assessments?*

Surveillance systems are largely designed and implemented at the local scale (country, state, etc) to support local public health decision making, but respiratory viruses exhibit global transmission patterns. Monitoring the transmission dynamics in other regions can often help predict future local dynamics. For seasonal influenza, epidemic dynamics during one hemisphere's winter can sometimes be suggestive of the future epidemic dynamics during the other hemisphere's winter (118); though there can still be much variation in epidemic dynamics even within a single hemisphere's flu season (119). During the SARS-CoV-2 pandemic, data from regions with high infection rates were relied on to inform risk assessment and scenario planning in regions with very low levels of infection or which had achieved local (temporary) elimination (120,121). When a novel variant emerges, detailed epidemic data can only be obtained in regions where there is a high degree of circulation. Analysis based on epidemic data obtained from other regions can be crucial for informing local public health strategies, especially for countries with common profiles. This may also benefit low-or middle-income countries since they may have insufficient resources to collect local data (21) (though there are some initiatives aimed at providing resources to such countries (122)). Sharing local surveillance data, and data analyses, between locales would be mutually beneficial, but there are open challenges; though some countries have shared enormous amounts of data both officially (*e.g.* through the WHO) and via country-specific public dashboards, there are others from which data has been minimal. Additionally, even when data are widely shared there can be international differences in data acquisition, definitions and reporting standards which make comparisons difficult. Successful examples exist such as GISAID (123), which has allowed global genomic sequence data to be shared widely; though there remain strong geographic biases in the sequences uploaded with North America and Europe over-represented (77,124).

*5.2 How can global genomic sampling procedures be optimised to improve vaccine strain selection?*

Vaccine strain compositions need to be updated regularly to manage the emergence of new antigenically distinct variants (SARS-CoV-2) / strains (influenza). Influenza vaccine compositions are reviewed bi-annually; SARS-CoV-2 vaccine compositions will likely also be reviewed frequently. The process by which influenza vaccine composition is selected is multifaceted and complex, and even after an appropriate strain is selected for inclusion the manufacturing process can take a significant amount of time. For this reason, vaccine composition recommendations are based on predictions of the dominant strains at some point in the future (77). Improving these predictions will improve future vaccine effectiveness and health outcomes.

Variants/strains can emerge anywhere. Seasonal influenza strains have been thought to predominantly be seeded from South/South-East Asia (125,126), though there is no guarantee that future strains will originate there. It is still not known what the geographic distribution of SARS-CoV-2 variants will be after an evolutionary equilibrium is reached, but in the first two years of the pandemic variants of concern were first detected in the UK (Alpha) (82), South Africa (Beta, Omicron) (127,128) and India (Delta) (129).

Predictions of future dominant variants/strains could be improved if they were detected and characterised earlier. More strategic and standardised global genomic sampling would decrease the time between emergence and detection, and potentially be more cost-effective. In the past there have been large geographic and temporal biases in sampling for influenza (77) and SARS-CoV-2 (75,124). However, as we do not know where in the world a new variant/strain will emerge, more geographically distributed sampling rates would improve global epidemiological insights and could be more cost-effective. In the future, geographic sampling distributions could be targeted to reflect any geographic heterogeneities in variant/strain emergence.

*5.3 How can gaps in global surveillance be handled?*

Complementing initiatives to improve within-country surveillance — such as increasing global coverage of existing sentinel surveillance systems and expanding them to cover multiple respiratory pathogens — methods have been proposed for leveraging traveller datasets (130). During the SARS-CoV-2 pandemic, international travellers were often required to undergo pre- or post-flight testing, as a method of reducing imported cases. The results from these tests were often informative for global surveillance; a country's local transmission dynamics could be inferred from the test results of people travelling out of the country (131). Genomic sequencing of these tests could further inform the dynamics of variants within the country. Testing requirements for international travellers have now been scaled back in many countries based on local public health surveillance and response requirements. While well justified from that perspective, it has also removed a valuable tool for global surveillance. It is not clear if there is a suitable surveillance mechanism that could perform this role during the post-pandemic period, and for viruses other than SARS-CoV-2. Testing and sequencing even a small percentage of travellers (for a specific respiratory pathogen) entering a few major airports would likely provide great insight into the pathogen's global patterns of prevalence, circulation and evolution. A surveillance system designed for routine testing of frequent international travellers, such as airline crews, could be one possible approach. Another possible approach that has been considered is testing the wastewater from international flights (132,133).

## 6. Conclusion

As the world transitions out of the acute phase of the COVID-19 pandemic, international and national surveillance systems are moving towards integrated models of surveillance for SARS-CoV-2, influenza, and other viral respiratory pathogens. This integration requires a re-evaluation of the public health objectives of surveillance, and consideration of how both pre-pandemic practices and new approaches adopted during the COVID-19 pandemic can best support those objectives. We have highlighted how different surveillance practices, previously applied to influenza and or COVID-19, contribute to specific areas of epidemiological analysis and insight. We have identified challenges associated with respiratory virus surveillance, many of which relate to the monitoring of individual viral respiratory pathogens, while others are specific to the development of integrated models of surveillance. Open questions remain on the design of integrated models of surveillance, including a need to further optimise individual components, and identify synergies and redundancies across them. Furthermore, the cost-effectiveness and feasibility of surveillance approaches in interpandemic and pandemic situations requires investigation; as does the transferability of information across countries and regions.

Our insights can assist surveillance planners as they assess the public health value and costs of various surveillance practices against the objectives of integrated models of surveillance for SARS-CoV-2, influenza, and other viral respiratory pathogens, supporting interpandemic surveillance and preparedness for the next pandemic.

**Author declarations**

The authors alone are responsible for the views expressed in this article and they do not necessarily represent the views, decisions or policies of the institutions with which they are affiliated.